\documentclass[sigconf, nonacm]{acmart}
\renewcommand\footnotetextcopyrightpermission[1]{} 
\AtBeginDocument{
  }

\acmConference[CHI '25]{The ACM CHI Conference on Human Factors in Computing Systems}{April 26--May 1, 2025}{Yokohama, Japan}

\usepackage{threeparttable}
\usepackage{fancyhdr}
\usepackage{graphicx}
\usepackage{xcolor}
\usepackage{colortbl}
\usepackage{url}
\usepackage{subcaption}
\usepackage{booktabs}
\usepackage{tabularx}
\usepackage{caption}
\usepackage{hyperref}
\usepackage{enumitem}
\usepackage{soul}
\usepackage{float}
\usepackage[inkscapelatex=false]{svg}
\usepackage{cleveref}
\begin{document}
\title{Interactive visualizations for adolescents to understand and challenge algorithmic profiling in online platforms}

\author{Yui Kondo}
\affiliation{%
  \institution{Oxford Internet Institute, University of Oxford}
  \city{Oxford}
  \country{UK}}

\author{Kevin Dunnell}
\affiliation{%
  \institution{MIT Media Lab, Massachusetts Institute of Technology}
  \city{Cambridge}
  \country{USA}}

\author{Isobel Voysey}
\affiliation{%
  \institution{Department of Computer Science, University of Oxford}
  \city{Oxford}
  \country{UK}}

\author{Qing Hu}
\affiliation{%
  \institution{School of Design, Carnegie Mellon University}
  \state{Pennsylvania}
  \country{USA}}

\author{Victoria Paesano}
\affiliation{%
  \institution{School of Computer Science, Massachusetts Institute of Technology}
  \city{Cambridge}
  \country{USA}}

\author{Phi H Nguyen}
\affiliation{%
  \institution{School of Computer Science, Carnegie Mellon University}
  \state{Pennsylvania}
  \country{USA}}

\author{Qing Xiao}
\affiliation{%
 \institution{Human-Computer Interaction
Institute, Carnegie Mellon University}
 \state{Pennsylvania}
 \country{USA}}

\author{Jun Zhao}
\affiliation{%
  \institution{Department of Computer Science, University of Oxford}
  \city{Oxford}
  \country{UK}
}

\author{Luc Rocher}
\affiliation{%
  \institution{Oxford Internet Institute, University of Oxford}
  \city{Oxford}
  \country{UK}}

\renewcommand{\shortauthors}{Kondo et al.}

\begin{abstract}
Social media platforms regularly track, aggregate, and monetize adolescents’ data, yet provide them with little visibility or agency over how algorithms construct their digital identities and make inferences about them. We introduce \textit{Algorithmic Mirror}, an interactive visualization tool that transforms opaque profiling practices into explorable landscapes of personal data. It uniquely leverages adolescents’ real digital footprints across YouTube, TikTok, and Netflix, to provide situated, personalized insights into datafication over time. In our study with 27 participants (ages 12–16), we show how engaging with their own data enabled adolescents to uncover the scale and persistence of data collection, recognize cross-platform profiling, and critically reflect algorithmic categorizations of their interests. These findings highlight how identity is a powerful motivator for adolescents’ desire for greater digital agency, underscoring the need for platforms and policymakers to move toward structural reforms that guarantee children better transparency and the agency to influence their online experiences.
\end{abstract}



\begin{teaserfigure}
  \includegraphics[width=\textwidth]{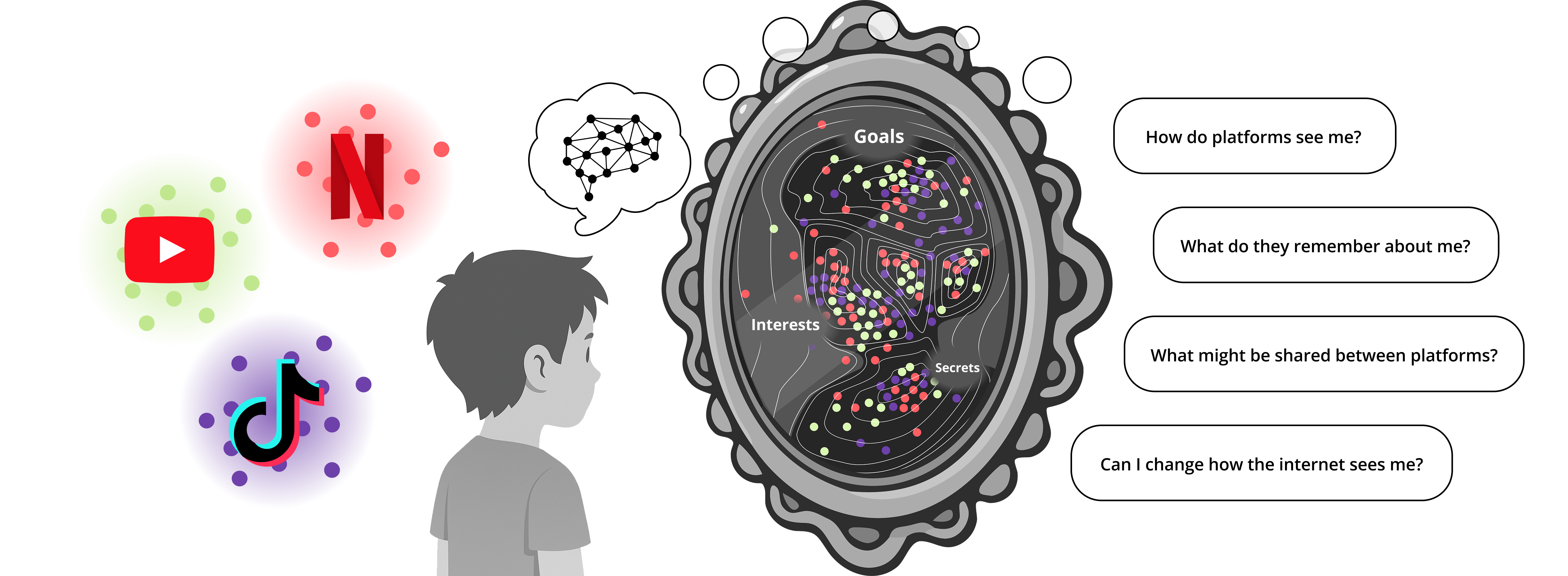}
  \label{fig:teasertest}
  \caption{Conceptual illustration showing how platforms construct a teen's profiles from data collected across services such as YouTube, Netflix, and TikTok, visualized through a mirror that reflects algorithmic interpretations of their goals and interests.}
\end{teaserfigure}

\maketitle

\section{Introduction}

Social media platforms have the power to radically transform how adolescents portray themselves and connect with each other~\cite{zizi_papacharissi_networked_2011,boyd2013connected}, often at the cost of their privacy and autonomy. Platforms offer opportunities for teens to practice autonomy and independence outside their family and the public eye, forming their identities, exploring their values, and establishing behavioral patterns that will persist into adulthood. On the other hand, platforms mediate adolescents' experience online, converting their everyday interactions, such as clicks, scrolls, and likes, into data for algorithmic analysis. This process, known as datafication~\cite{van2014datafication,wang2022don}, enables platforms to create a detailed digital representation (or `mirror') of their users' preferences, providing each user with unique, personalized content recommendations to maximize screen time and engagement~\cite{raffoul2023social}.

Researchers have shown that the algorithmically mediated design of social media platforms strongly affects children, particularly adolescents, with important consequences on their mental health, attention span, and overall digital well-being~\cite{livingstone2017young, 5Rights2023, santos2023associations}. Adolescence represents a crucial window of vulnerability to social media~\cite{orben2022windows}. During this period, teens are especially sensitive to immediate feedback and social rewards, while their impulse control and critical thinking are still maturing according to developmental psychology~\cite{piaget1972development,kuhn1999developmental,blakemore_inventing_2018,duell_positive_2019,crone_understanding_2012}. This creates a developmental imbalance that heightens susceptibility to persuasive, personalized feeds~\cite{steinberg2005cognitive, casey2008adolescent,romer_executive_2009,shulman_dual_2016,telzer_dopaminergic_2016}. Algorithmic systems that mediate these experiences can shape not only adolescents' behavior in the moment but also leave lasting impacts on their development~\cite{park2025towards,qi2024digital,manning2021framework}, and evolving sense of self~\cite{granic2020beyond,wood2016digital,handayani2020impact}. 

Adolescents are losing control of their digital lives but lack the tools to make informed use of social media~\cite{schaper2023five,veldhuis2025critical}. Current policies and interventions have often focused on restricting content or limiting access, rather than on empowering younger users~\cite{hertog2024data, wang2021protection}. Adolescents are left behind a `one-way mirror', with algorithms constructing detailed profiles to shape their online experiences, while being excluded from understanding these digital representations. Without support to make sense of opaque datafication practices and inferences applied to them, adolescents struggle with critical reflection, underestimate the extent of profiling, and experience a diminished sense of agency over their digital identities~\cite{livingstone2014developing,livingstone2017young, granic2020beyond}. 

To expose this asymmetric relationship between platforms and adolescents, researchers have explored a range of tools and frameworks aimed at fostering adolescents' datafication awareness and critical algorithmic literacy~\cite{micallef2021fakey,chou2023misinformationgame,wang2024chaitok}. However, while these efforts have produced encouraging findings, most have been limited to a controlled setting through simulations, games, or synthetic data, disconnected from adolescents' lived experiences ~\cite{papert1991constructionism,dignazio_creative_2022}. As a result, adolescents continue to struggle with abstract datafication concepts, especially given the prevalence and scope of profiling and personalization across platforms and time~\cite{wang2024chaitok,hertog2024data}.

Here, we propose and test a novel interactive tool to support adolescents in reflecting on real social media usage across platforms and over time. This tool, Algorithmic Mirror, allows adolescents to visualize their watch histories from YouTube, Netflix, and TikTok, three platforms that collectively shape adolescents' media consumption worldwide. Our hypothesis is that visualizations based on real data, not simulations, can help adolescents better understand the digital identities that social media algorithms learn about them, and in turn, foster intrinsic motivation and autonomy to shape the content they want to consume online. To test this hypothesis, we conducted a two-part user study with 27 adolescents aged 12--16 through a qualitative method. Each participant requested their video viewing histories, ranging from 100 to 60,000 videos across the three platforms and up to five years of activity, and uploaded them to Algorithmic Mirror. We then conducted interviews, guiding them to explore how their digital identities have evolved, compare how different platforms perceive them, and understand the cumulative nature of datafication. Overall, we investigated four research questions:

\begin{itemize}
    \item[$\diamond$] RQ1: How do adolescents currently perceive data collection and inference on social media platforms?
    \item[$\diamond$] RQ2: How does Algorithmic Mirror affect adolescents' experience and their perception of datafication?
    \item[$\diamond$] RQ3: How does Algorithmic Mirror enable adolescents to reflect on how their online experiences are shaped by algorithmic inferences and datafication? 
    \item[$\diamond$] RQ4: What are adolescents' needs for tools that help them critically understand and navigate datafication, and what are the barriers they face in exercising data autonomy?
\end{itemize}

We show that Algorithmic Mirror effectively makes the scale and scope of datafication across platforms and over time directly visible to young people. Algorithmic Mirror enables adolescents to make personal connections with their lived experiences and fosters self-reflection through personalized visualizations from their own data. Finally, our findings identify how to best help adolescents develop critical awareness of algorithmic systems and datafication, with important insights for researchers, educators, policymakers, and platform developers.

\section{Related Work}
\subsection{Datafication Around Young People}

Datafication refers to the translation of increasingly diverse aspects of social life into quantifiable digital traces that can be stored, analyzed, and monetized~\cite{mejias2019datafication,van2014datafication}. Mainstream social media platforms such as Facebook, Instagram, TikTok, and YouTube, as well as online platforms such as Amazon Prime and Netflix, routinely collect vast amounts of personal data ranging from explicit content (profiles, conversations, images), behavioral metrics (likes, purchases, engagement patterns), to metadata~\cite{raffoul2023social,han2022dare}. These data are processed by algorithms to infer hidden attributes and are aggregated to construct comprehensive user profiles for targeted advertising, personalization, and risk assessment~\cite{truong2024account,treves2025viki}.

Children are deeply embedded in such a data-based economy, giving rise to what is termed `datafied childhood'~\cite{mascheroni2020datafied}. From birth, children’s digital identities are shaped by persistent data collection, aggregation, and inference, producing marketable profiles traded in opaque data economies, or surveillance capitalism~\cite{zuboff2019age}. This process not only facilitates sustained profiling but also enables manipulative design strategies, such as dark patterns~\cite{van2022don}, stealth advertising~\cite{rozendaal2023children}, and gamified inducements~\cite{meyer_advertising_2019}, which encourage prolonged screen engagement, oversharing, and impulsive spending~\cite{aap2020digital}. Social media platforms are a prevalent actor in exercising datafication, tracking children and young people's online interactions, preferences, and peer networks to fuel their key business revenues through targeted advertising and algorithmic recommendations. As shown by Raffoul et al.~\cite{raffoul2023social}, ``approximately 30-40\% of the advertising revenue generated from three leading social media platforms is attributable to young people'', which amounts to nearly \$11 billion US dollars per year. This section outlines the pervasiveness and harms of datafication for adolescents and our work focuses on supporting minors in developing their own capacities to navigate and resist such practices.


\subsection{Adolescents’ Data Autonomy and Critical Algorithmic Literacy}
Pervasive datafication practices not only harm children's online safety and digital well-being but also undermine their autonomy to control their data, attention and behavior~\cite{sahebi2022social, peterson2020negotiated, furnham2019automation}, with lasting impacts on their digital resilience and ability to make informed choices as they develop into adulthood~\cite{park2025towards,qi2024digital,manning2021framework}. In response to these concerns, there's a growing consensus on the importance of fostering children's autonomy in the digital space, which includes developing their understanding, values, self-determination, and self-identity~\cite{danby2017technologies,vidal2020early,wang2023child}. Autonomy refers to a person's effective capacity for self-governance and acting in accordance with their own beliefs, values, and motivations ~\cite{darwall_value_2006, friedman_autonomy_1986, klenk_autonomy_2019}. This is highly contextualized by the unique features of adolescence, a formative stage in which both inner values and self-regulatory capacities are rapidly developing. Developmental psychology shows that adolescents' reward sensitivity peaks while their self-regulatory skills are still maturing, making them more vulnerable to persuasive and personalized influences~\cite{steinberg2005cognitive,steinberg2010dual,casey2008adolescent,romer_executive_2009,shulman_dual_2016,telzer_dopaminergic_2016}. At the same time, their capacity for abstract reasoning and critical thinking begins to consolidate~\cite{piaget1972development,kuhn1999developmental,blakemore_inventing_2018,duell_positive_2019,somerville_charting_2017,crone_understanding_2012}, making this life stage crucial for cultivating autonomy.

Research has shown fundamental gaps in adolescents' understanding of datafication, which further impairs their ability to exercise their autonomy over their data, attention and behavior~\cite{wang2022don,sun2021they,kumar2023understanding}. For example, research finds that adolescents can often articulate interpersonal privacy tactics, such as adjusting account privacy settings, and many recognize that platforms monetize their data~\cite{goray2022youths,agha2024tricky}. Yet, understandably, they struggle to comprehend complex institutional data flows, such as cross-platform data tracking, profiling, and inferential analytics, limiting their capacity to foresee collective or long-term risks (e.g., sensitive attribute inference, reputational trail)~\cite{livingstone2018european,smahel2020eukidsonline,wang2024chaitok,sun2024why,reyes2021money}. These gaps reveal that, while adolescents can make localized, experience-based choices, their developmental stage leaves them underprepared to make genuinely informed decisions about complex socio-technical systems~\cite{badillo2019stranger,chowdhury2023co,dempsey2022children}.

To achieve data autonomy, adolescents need to understand algorithms, predictive inferences, and their long-term impact~\cite{williamson2022education,zuboff2019age}. Scholars argue for a critical shift from generic `digital literacy' to critical data literacy, which makes visible how data are generated, interpreted, and used to shape opportunities~\cite{pangrazio2016reconceptualising}. Critical data literacy situates data within power, context, and rights, moving beyond safety tips towards interpretive, ethical, and tactical capacities~\cite{pangrazio2019personal, kafai2020theory, morales-navarro_conceptualizing_2023}. Prior research has demonstrated how this framework encourages adolescents to question invisible dataflows, understanding how platforms collect, analyze, and serve data beyond basic technical know-how~\cite{wang2022don}.  Furthermore, such skills are also crucial for adolescents to critically examine how their `digital self' is constructed, enabling them to resist or reshape algorithmic representations and demand better data autonomy~\cite{wang2024chaitok}. However, there remains a gap in making cross-platform data sharing and inference chains both visible and actionable for young people.

\subsection{Tools for Datafication Awareness and Reflection}

HCI researchers have tested a range of tools and frameworks to raise datafication awareness by fostering critical algorithmic literacy, often limited by their abstract nature. Game-based simulations such as Fakey~\cite{micallef2021fakey}, which mimics a social media news feed to let players practice evaluating credibility, have shown promise in improving skepticism toward low-credibility sources. Interactive simulations such as The Algorithm~\cite{elkattan2023algorithm}, The (Mis)Information Game~\cite{chou2023misinformationgame}, and CHAITok~\cite{wang2024chaitok} similarly use controlled environments to illustrate how feeds and sharing decisions are shaped by algorithmic logic. 

{\color{blue}
}


However, studies have highlighted a limitation of these controlled approaches: when learning is decontextualized from the learner's own digital life, the resulting knowledge often remains abstract and difficult to transfer into practice~\cite{wang2024chaitok}, as online experiences become dissociated from memory~\cite{baughan2022dissociation}. Recent work in data literacy reinforces this point, showing that educational tools create significant barriers when they fail to engage with learners’ personal experiences and existing mental models~\cite{dignazio_creative_2022, dangol_constructionist_2023}. 
In a similar way, reflective informatics (the body of work in HCI focused on designing for reflection~\cite{baumer_reflective_2015}) emphasizes linking to a learner's real experiences to facilitate meaningful reflection~\cite{slovak_reflective_2017}.
These findings underscore the need for human-centered mechanisms of algorithmic transparency~\cite{ehsan_human-centered_2022, luria_co-design_2023}. Motivated by this, we ground our work in participants’ personal data, which provides the authenticity and situational relevance often lacking in current approaches.

This approach has also been explored in previous tools that expose learners to their real personal data and promise more authentic engagement. FeedVis revealed how Facebook's News Feed filtered posts compared to the uncurated stream~\cite{eslami2015feedvis}, while others allowed users to visualize their own `filter bubbles'~\cite{nagulendra2014filterbubble}. Projects such as Gobo~\cite{bhargava2019gobo} and Lightbeam~\cite{mozilla2013lightbeam} extend the scope by integrating data across multiple services. Conducting research with real-world data introduces privacy and complexity challenges, yet makes insights more meaningful and transferable than abstract simulations~\cite{dasgupta2021designing}.
However, tools designed to support adolescents to grow more aware of datafication and reflect on its prevalence and impact, like FamiData Hub~\cite{Wang2025FamiDataHubb}, KOALA Hero~\cite{wang_koala_2024} and CHAITok~\cite{wangCHAITokProofofConceptSystem2024}, currently do not use adolescents' real-world data.

Furthermore, existing tools for datafication awareness rarely capture both spatial and temporal data dimensions. Spatially, datafication functions through extensive cross-platform data sharing and profile aggregation, yet early systems such as FeedVis~\cite{eslami2015feedvis} provide only single-platform snapshots. While later tools like Gobo and Lightbeam offer cross-platform views, they remain largely static or momentary, failing to capture how platforms accumulate comprehensive behavioral profiles over extended periods ~\cite{bhargava2019gobo, mozilla2013lightbeam}. Temporally, datafication is inherently accumulative: platforms build detailed profiles over months and years of user activity. Existing longitudinal visualizations such as Themail~\cite{viegas2006themail} and Visits~\cite{thudt2016visits} emphasize temporal unfolding but remain limited to single domains, missing the cross-platform nature of contemporary data collection. Our work addresses these limitations by integrating both spatial and temporal dimensions of datafication. We combine cross-platform data visualization with a longitudinal analysis spanning years of user activity. This approach enables situated, personalized reflection on how datafication accumulates and shapes user experiences over time, bridging the gap between abstract knowledge and lived digital experience, which influences users over years.

\section{Design}\label{sec:method}

\begin{figure*}[t]
    \centering
    \includegraphics[width=1.0\textwidth]{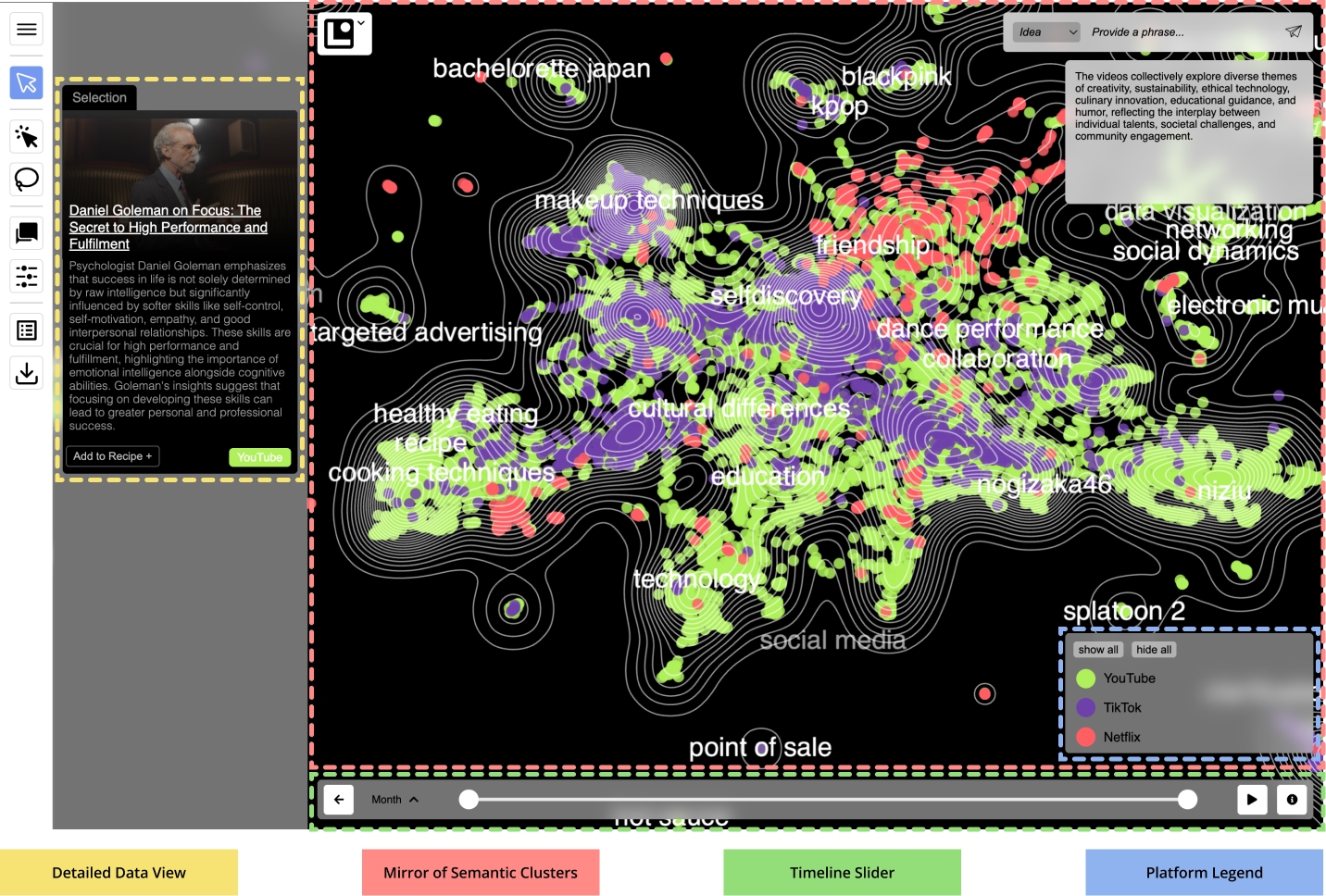}
    \caption{Algorithmic Mirror Interface. The four main components of the interface are highlighted with dashed lines.}
    \label{fig:data_mirror}
\end{figure*}

\subsection{System Design and Development}
Our primary design motivation for Algorithmic Mirror was to create personalized visualizations of adolescents’ social-media watch histories that support reflection on digital identity and algorithmic perception. Prior work on “filter bubbles” (algorithmically reinforced information enclaves), such as Roy’s The Internet Needs You-Are-Here Maps~\cite{GillaniRoy2022} and the Center for Constructive Communication’s Me, My Echo Chamber, and I project~\cite{gillaniMeMyEcho2018}, demonstrates the value of revealing clustered information environments. Research on dynamically constructing such visualizations based on semantic embeddings led to Latent Lab~\cite{dunnellLatentLabExploration2023}, a publicly available system for transforming high-dimensional, unstructured data into navigable 2D semantic maps with arbitrary user-provided corpora. Algorithmic Mirror extends Latent Lab, introducing a module for preprocessing raw platform exports and integrating the existing sentence-embedding models, LLM-based theme extraction, and dimensionality-reduction techniques into a unified pipeline. The result is a user-friendly tool that makes exploration of digital identities across both spatial (cross-platform) and temporal (multi-year) dimensions accessible to adolescents.

Algorithmic Mirror is designed with four key components: personal data uploading (\Cref{subsubsec:personal-data-uploading}), semantic clustering (\Cref{subsubsec:semantic-clustering}); temporal exploration (\Cref{subsubsec:temporal-analysis}); cross-platform data integration (\Cref{subsubsec:cross-platform-data-integration}).

\subsubsection{Personal Data Uploading}
\label{subsubsec:personal-data-uploading}

\begin{figure*}[t]
    \centering
    \includegraphics[width=1.0\textwidth]{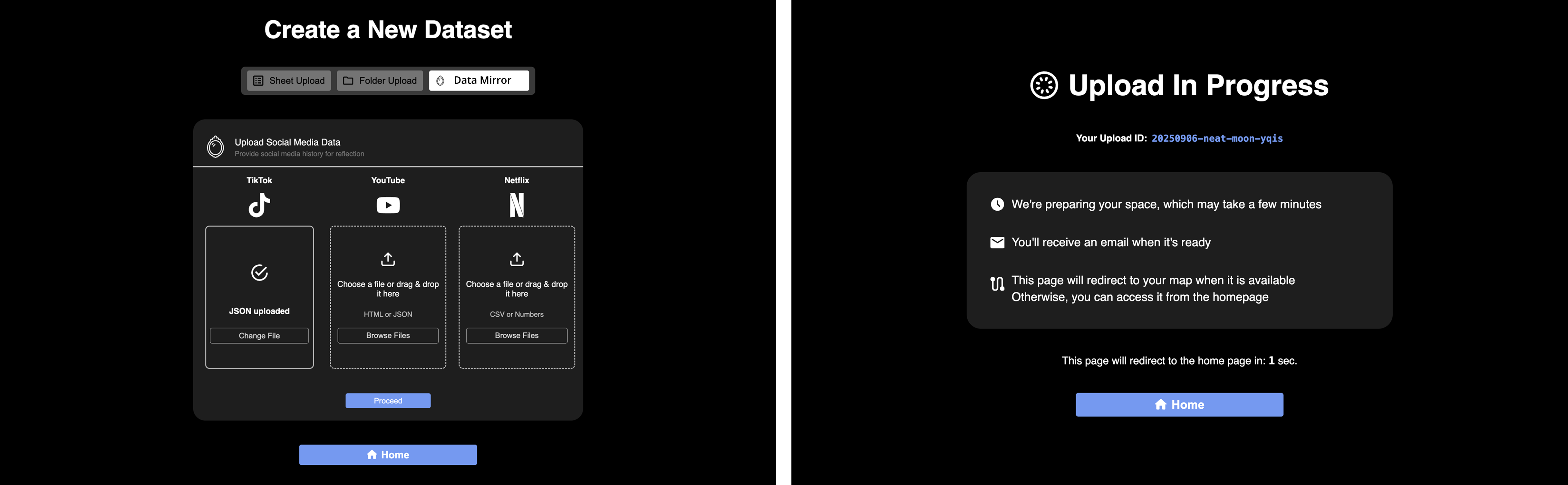}
    \caption{Algorithmic Mirror upload user flow. Users can drag and drop exported watch histories from three supported platforms. Once uploaded, the automated processing is initiated, and users receive an email when the processing is complete.}
    \label{fig:user_data_upload}
\end{figure*}

\Cref{fig:user_data_upload} shows the user-friendly data upload component we developed, with a drag-and-drop interface that accepts platform-specific exports. Since YouTube, TikTok, and Netflix do not expose viewing-history APIs, participants obtained their data through each platform’s GDPR subject access request tools. Our upload module processes these raw exports using platform-specific pipelines, converting heterogeneous file formats into a unified structure for visualization.

\begin{figure*}
    \centering
    \includegraphics[width=0.25\textwidth]{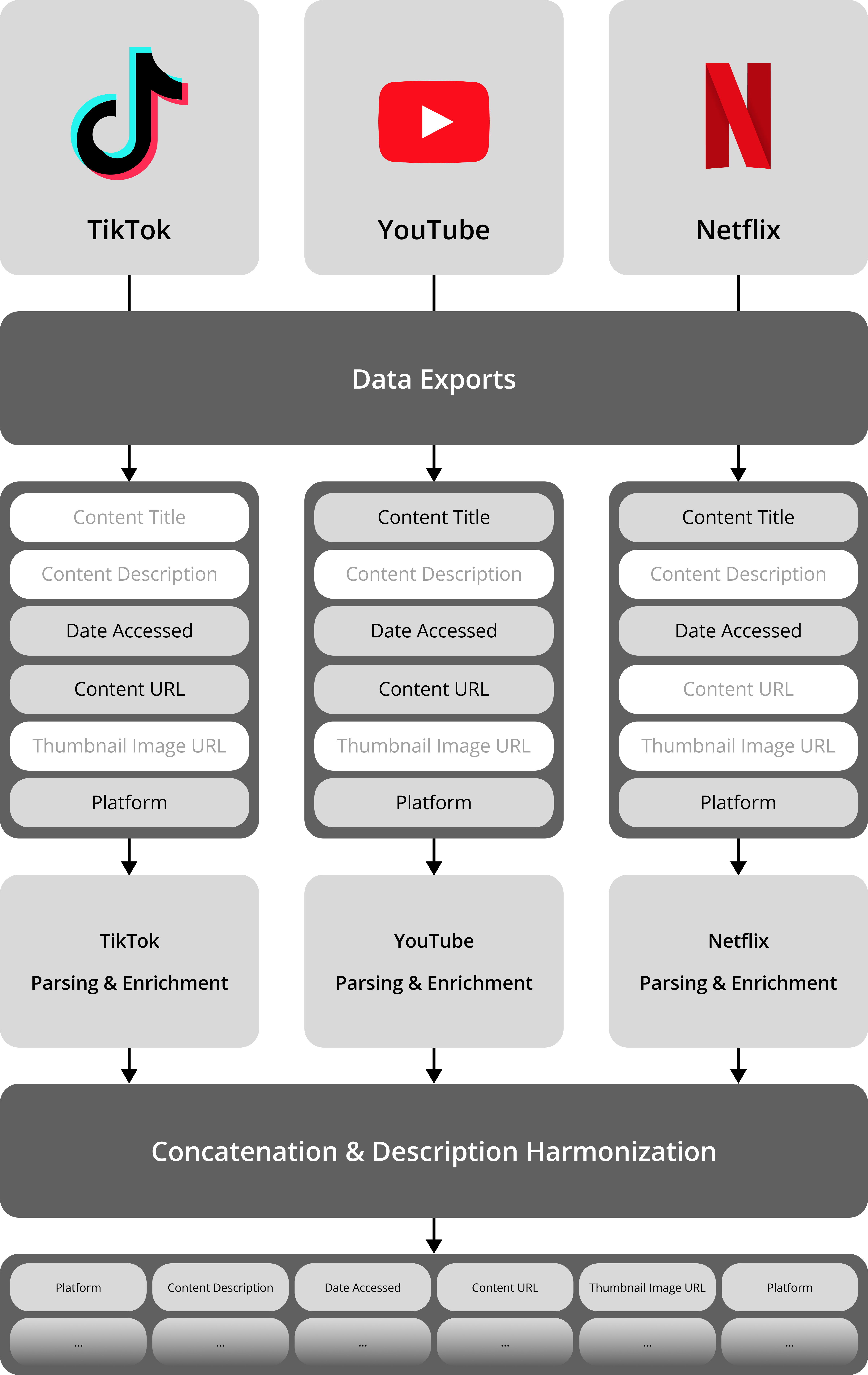}
    \caption{Algorithmic Mirror pre-processing pipeline. User watch history exports contain sparse and platform-specific data. The pre-processing pipeline enriches these exports by filling in missing information, concatenating disparate sources into a unified dataset, and harmonizing heterogeneous content descriptions for each video before passing them to the embedding and topic extraction modules.}
    \label{fig:pre_processing_pipeline}
\end{figure*}

Figure~\ref{fig:pre_processing_pipeline} summarizes this workflow and the metadata fields that each platform provides (black text) versus omits (gray text). In all cases, the exported data lacked complete content descriptions and therefore required additional enrichment. For TikTok, viewing-history files contain URLs and timestamps but no titles or descriptions, so each URL was resolved to retrieve the creator’s title and caption. YouTube exports include URLs, titles, and timestamps; these were augmented by retrieving automatic transcripts when available through YouTube’s API. Netflix exports were drawn from a child’s profile within a shared family account and provided only titles and timestamps; each title was used to query The Movie Database (TMDB) API\footnote{\url{https://www.themoviedb.org/}} to obtain content descriptions. Since the resulting descriptions varied widely in syntax and emphasis (e.g., TikTok captions dominated by hashtags, YouTube descriptions containing promotional links, Netflix synopses oriented toward narrative arc), an LLM (GPT-4o-mini ~\cite{OpenAI_GPT4oMini_2024}) was used to harmonize them to comparable length and a consistent focus on the core content being described. The resulting standardized descriptions were then passed to the embedding and topic-extraction modules for further processing.

\begin{figure*}[t]
    \centering
    \includegraphics[width=1.0\textwidth]{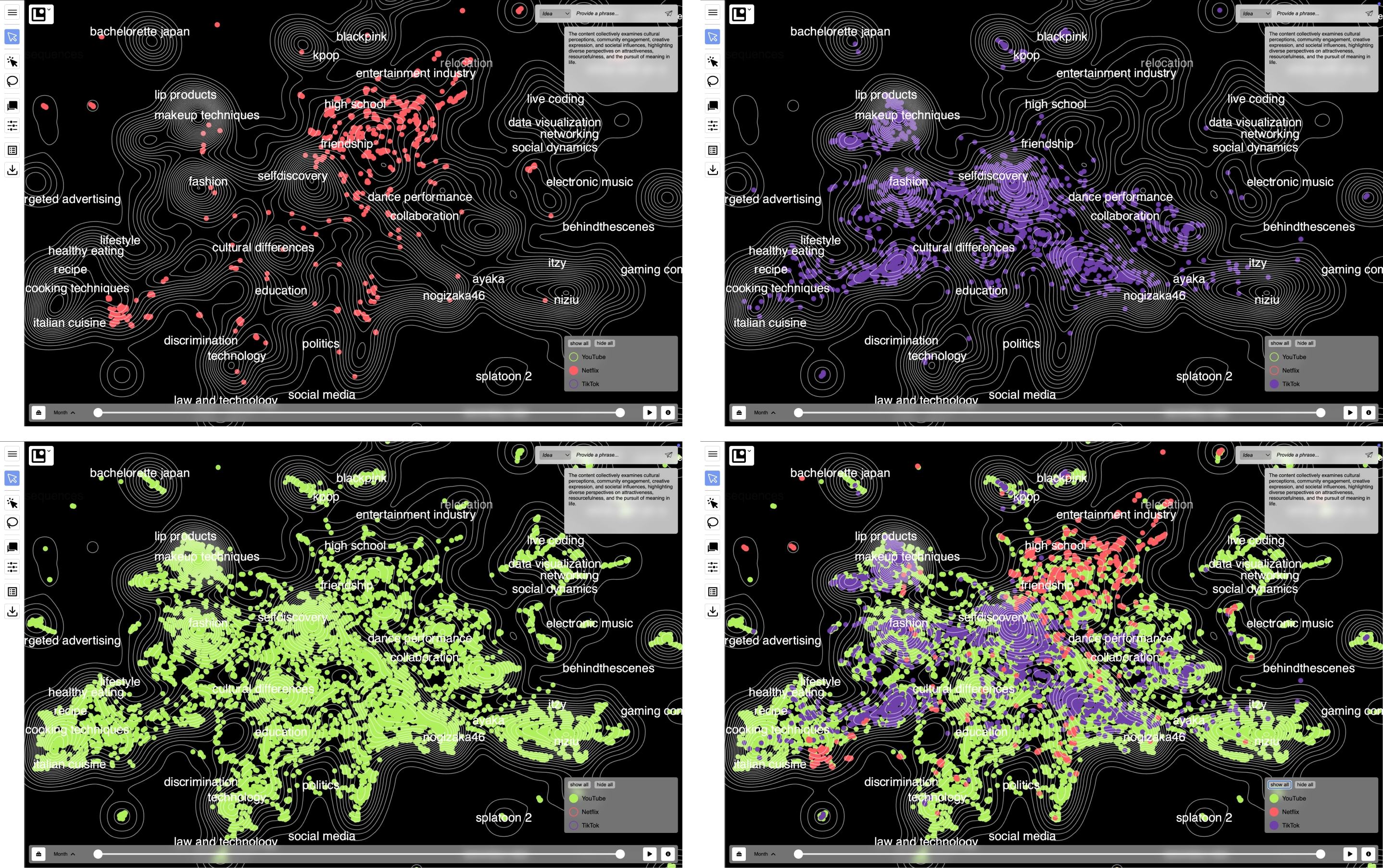}
    \caption{Cross-Platform Data Integration. The top-left shows Netflix data, the top-right TikTok data, and the bottom-left YouTube data. The bottom-right overlays all three, where intermingled clusters reveal semantically similar content descriptions across platforms, illustrating how a persona is constructed across different services.}
    \label{fig:datafication}
\end{figure*}

\subsubsection{Semantic Clustering for Algorithmic Awareness}
\label{subsubsec:semantic-clustering}
The resulting `mirror' displays a map of video thumbnails representing the user's viewing history, with each video color-coded by its originating platform (see \Cref{fig:data_mirror}). When zoomed out, individual videos appear as colored dots. Dot position reflects semantic similarity--videos with related themes cluster together--while color indicates which platform supplied the content. We automatically generate these semantic clusters by converting video descriptions into numerical vector representations using OpenAI's text-embedding-ada-002 model \cite{openai2022ada002}, then applying Universal Manifold Approximation and Projection (UMAP) \cite{McInnes2018} to reduce these high-dimensional vectors to 2D coordinates while preserving the local and global semantic relationships in 2D. This approach resembles the semantic relationship bucketing that platforms use to infer user preferences and drive content recommendations.

Topics are ranked by frequency. Commonly referenced topics appear prominently, while less frequent topics become visible only upon zooming (see Zoom and Pan feature diagram, \Cref{fig:zoom_and_pan_and_temporal_evolution}). This hierarchical organization of topics aims to give users a sense of how algorithms prioritize and weigh different aspects of their own behavior. It helps users understand what platforms consider to be their main interests. These topic labels are positioned near semantically similar videos to demonstrate the associative logic underlying recommendation systems. Users can interact with topic labels to view all associated individual videos, enabling them to audit and reflect on the accuracy of algorithmic categorizations of their viewing patterns.

\subsubsection{Temporal Analysis}
\label{subsubsec:temporal-analysis}
\begin{figure*}
    \centering
    \includegraphics[width=1.00\textwidth]{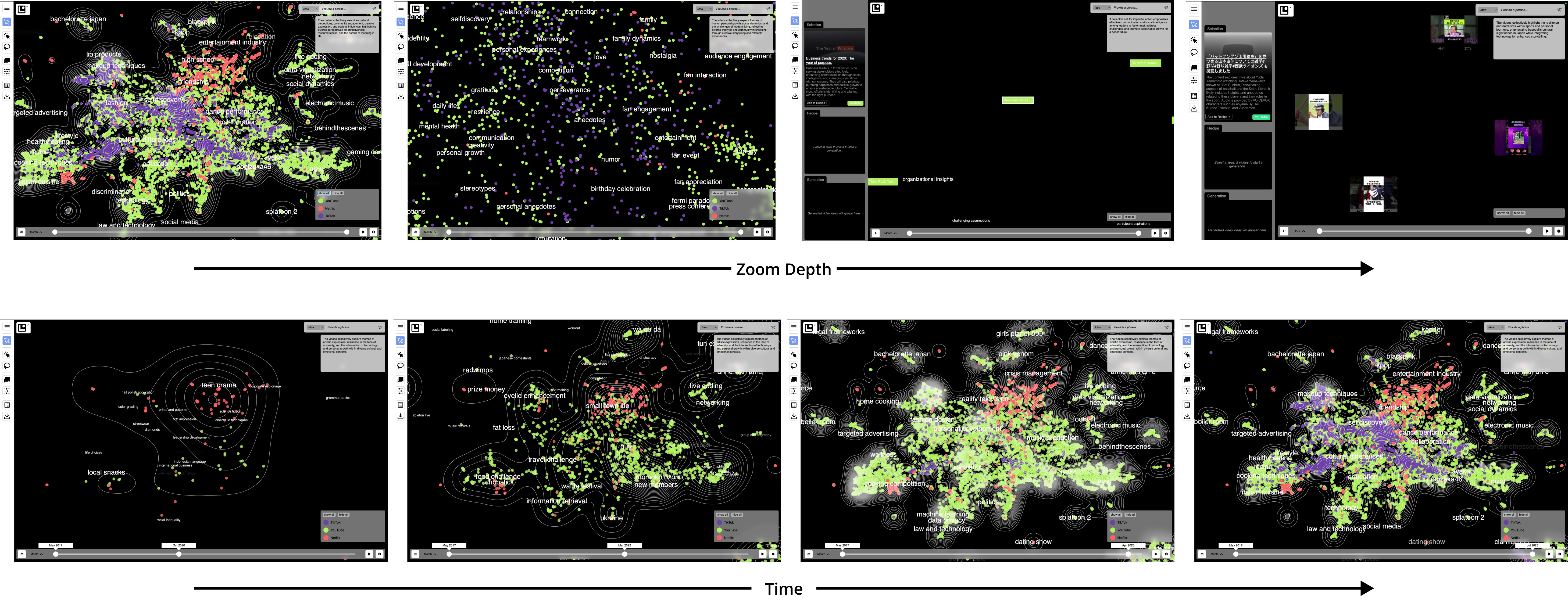}
    \caption{Top: Zoom and Pan feature. Users can pan across the Mirror to explore different regions of content. Zooming in reveals less frequent subtopics within higher-level themes. At high zoom levels, contour lines indicate dense clusters, which fade as individual items appear—first by title, and eventually by thumbnail. The dot color indicates the platform where each video was watched.
    Bottom: Timeline Slider feature. Users can drag the slider to explore how their viewing content evolves over time, or click play to watch the progression unfold automatically. The starting point can be adjusted to focus on specific periods and reveal the most common themes within that time frame.}
    \label{fig:zoom_and_pan_and_temporal_evolution}
\end{figure*}

Algorithmic Mirror incorporates an interactive timeline feature that enables cross-platform temporal analysis of viewing patterns (see bottom of   \Cref{fig:zoom_and_pan_and_temporal_evolution}). Users can observe how their content landscape evolves over time, watching topics change in relevance in the dynamic visualization. The interface is designed to help users reflect on their watch interests over time. For example, some users may see how trending content bubbles to prominence before quickly disappearing, while foundational interests demonstrate steady growth and gradually shift toward central positions in the visualization space.

\subsubsection{Cross-Platform Data Integration}
\label{subsubsec:cross-platform-data-integration}
A key contribution of Algorithmic Mirror is exposing the datafication of users across platforms in a single, homogenized view. This requires clustering similar content from different platforms in the same location on the visualization. Initial implementations revealed a clustering bias where content from each platforms formed distinct, separate clusters (i.e., one cluster for YouTube, one for TikTok, and one for Netflix). This segregation resulted from syntactic variations in video content descriptions across platforms: Netflix descriptions typically contain clean content synopses, TikTok captions are brief with extensive emoji and hashtag usage, while YouTube descriptions often include timestamps, external links, and promotional content.

To address platform heterogeneity, we implemented a homogenization preprocessing step (the final preprocessing step in \Cref{fig:pre_processing_pipeline}). Raw video descriptions are processed with OpenAI's GPT-4o-mini model~\cite{OpenAI_GPT4oMini_2024} using prompts designed to generate short summaries that focus exclusively on content, while filtering out promotional material, hashtags, and external links. This approach improved the final visualization layout, achieving better content intermingling across platforms, as demonstrated in the cross-platform data integration shown in \Cref{fig:datafication}.

\subsection{Privacy Considerations}
Privacy protection is fundamental to Algorithmic Mirror's design, given the sensitive nature of personal viewing data. We implemented several safeguards throughout the data processing and storage pipeline, including encrypted, secure access for each study participant and storage limited to video metadata and embeddings, with no personally identifiable information (PII) from raw exports retained.

Our initial design considered richer personal data sources such as browser histories and LLM chat logs, but these contain substantially more PII and would amplify privacy risk when using a hosted embedding service. Because OpenAI retains input data for up to 30 days solely for abuse monitoring \cite{openai2024retention}, we adopted a data-minimization strategy: only non-identifying, content-level information (e.g., YouTube, Netflix, TikTok video titles and descriptions) was sent for processing. This choice also reflects a practical trade-off: delivering responsive performance for many simultaneous users requires peak computational capacity that is difficult to provision and maintain with purely local university servers, whereas hyperscale cloud infrastructure provides throughput necessary for real-time embedding at scale. This approach reduces exposure risk while still enabling the performance necessary to process large viewing histories. With dedicated compute, the system would migrate to fully local model deployment, eliminating external transmission entirely.

\section{User Study}
\label{sec:user-study}

We conducted 27 individual user study sessions in July-September 2025, post-IRB approval. Each study involved one to four participants and was led by at least two researchers via video conferencing using participants' personal devices with their own social media data. The study was designed to minimize data privacy concerns by allowing participants to maintain control of their personal accounts throughout the process. The complete study spanned approximately two weeks per participant and consisted of three main phases: a pre-study to filter participants, a data preparation phase, and the main study. The main study was designed to use a combination of open exploration and think aloud~\cite{eccles2017think} method, along with a closing semi-structured interview. In the `think aloud' model, study participants are encouraged to verbalize their thoughts, feelings, actions, and decisions while performing certain tasks. Research has shown that it is an effective method to uncover participant's cognitive process, identify any points of confusions, while at the same time allowing (particularly young) participants to openly explore and raise questions~\cite{van2018x}. We complemented this with guiding questions (see below) and observational notes. Notes were particularly crucial when the study involved more than one participant, allowing us to effectively interpret the audio transcripts during data analysis. As a result, we used a mixed method by combining participants' self-reporting experiences, think-aloud data, and researchers' observations. 

\begin{figure*}
    \centering
    \includegraphics[width=1\linewidth]{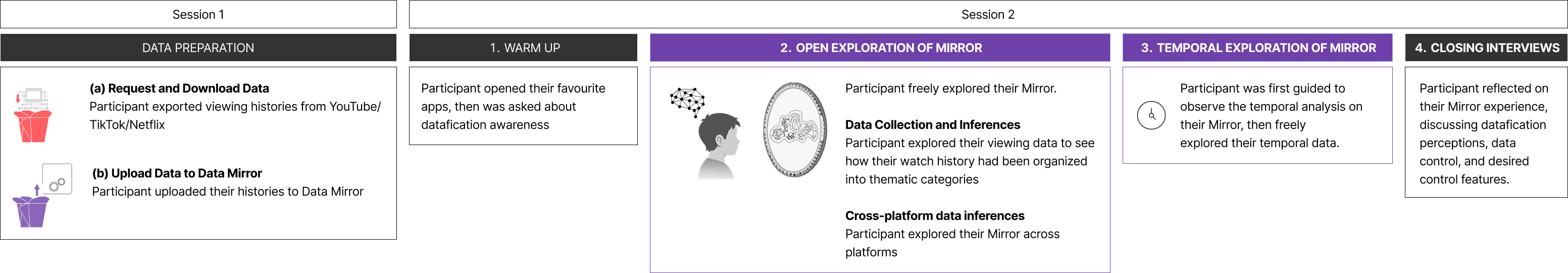}
    \caption{Study procedure}
    \label{fig:study procedure}
\end{figure*}

\subsection{Participants}

We recruited 27 children aged 12-16 through convenience and snowball sampling on social media and educational institutions, following CUREC Approved Procedure AP25 for non-invasive research with children recruited via organizations.  The participants were required to be active social media users (YouTube, Netflix, or TikTok).
We selected this age group for several reasons. 
Adolescents are heavy users of these platforms, many spending multiple hours on social media platforms each day~\cite{rothwell_how_2023}, and are at a developmental stage where they are highly susceptible to persuasive, personalized feeds~\cite{steinberg2005cognitive,steinberg2010dual,casey2008adolescent}. It is thus essential to assess the experiences of children at this stage regarding social media datafication, as they represent users who can make localized choices but may struggle with understanding complex institutional data flows, such as cross-platform tracking and inferential analytics~\cite{livingstone2018european,smahel2020eukidsonline}.
Among the 27 participants, two were 12, six were 13, nine were 14, eight were 15, and two were 16 years old (average of 14.1). Fourteen were girls and thirteen were boys. Participants were geographically distributed across Japan ($n=16$), the UK ($n=4$), Europe ($n=4$), and other countries, including Bahrain and the US ($n=3$). While the sample was skewed toward Japanese participants due to school recruitment availability, our analysis revealed that geographic differences had minimal impact on participants' responses (see \Cref{sec:limitations}).
The most common platform data uploaded by participants was YouTube ($n=22$), followed by Netflix ($n=11$) and TikTok ($n=5$). Nine participants uploaded data from two or more platforms (see \Cref{tab:demographics} for further details).

For consent procedure, we followed a dual consent and assent procedure. We obtained written consent from parents or guardians and assent from adolescents, ensuring they understood the study and confirming their willingness to participate. Both parents and children were informed that participation was voluntary and could be withdrawn at any time without consequences or affecting compensation. Participants could withdraw their data until 30 September 2025, by when we estimated analysis would be completed.

For compensation, each participating family received a £10 gift card for study preparation (downloading their data) and an additional £10 gift card for completing the main study session. 

\subsection{Group Formation}

Participants were grouped by age and gender to ensure developmentally appropriate peer interactions and comfort levels during discussions of potentially sensitive data visualizations. Group sizes varied based on recruitment availability and scheduling constraints. To further enhance participant comfort, we considered family relationships and school connections when forming groups. Where possible, we accommodated siblings or family members who wished to participate in the same session, and we grouped classmates together when they were recruited from the same school. These grouping strategies were particularly important given that participants would be opening and discussing their personalized data visualizations (the "mirror") in front of others, which could potentially be uncomfortable. The presence of familiar peers or family members helped create a supportive environment that encouraged open discussion while respecting individual privacy boundaries. During group discussions, participants controlled what they chose to share about their visualizations, and researchers emphasized that sharing was optional.

\subsection{Data Preparation}
Following a pre-survey, we set up a data preparation session for participants of suitable ages and with experience of relevant social media platforms.  As described in Section~\ref{sec:method}, the distinctive feature of Algorithmic Mirror is to allow study participants to gain a deep reflection of their online activities and datafication by examining their real social media data. Thus, we take the highest caution to protect all participants' data privacy by introducing a separate `data preparation' session. This session allows them sufficient support to request data from social media platforms and upload the data directly to our tool, Algorithmic Mirror, without exposing any of their private information to the researchers in charge of interviews.

During the data preparation session (up to 60 minutes), researchers provided each participant step-by-step guidance for downloading personal data from their chosen platforms (YouTube, Netflix, TikTok) using the platforms' official data export features. Following this, researchers provided email support to help participants securely upload their data to the Algorithmic Mirror platform, with all personal identifying information removed before upload. This preparatory phase is aligned with constructionist learning principles~\cite{papert1991constructionism}, which emphasize that learners construct knowledge most effectively when working with personally meaningful materials. By inviting participants to work with their own real digital footprints rather than abstract simulations, this approach enabled contextualized reflection during the subsequent activities where participants could construct their own understanding of digital behavior patterns. 

\subsection{Main Study}
Each session lasted approximately 60 minutes. When availability allowed, participants worked in age-grouped pairs/groups to share observations and collectively reflect on algorithmic influence on their social media usage. Participants were grouped by age to ensure developmentally appropriate peer interactions, with group sizes varying based on recruitment availability and scheduling constraints. 

\subsubsection{Warm Up (10 minutes)}
It is crucial for teens to feel comfortable working with adult researchers, and other peers in their age-paired group when applicable. We kicked off each session with warm-up activities, inviting participants to show us their favorite social media platforms, how they search for content, and what they think their interests are. This enabled us to gather initial insights about their understanding of datafication on social media platforms.

\subsubsection{Part 1: Open Exploration of Mirror (25 minutes)}
At the start, participants were introduced to their personalized Mirror visualization through a brief orientation. They were then encouraged to freely explore their Algorithmic Mirror while researchers observed and took notes. Participants were encouraged to think aloud during their exploration and ask questions about what they observed. The exploration was guided by a series of open-ended prompts designed to scaffold reflection on seeing their collected data (e.g. are you surprised by what you see) and examination of general data patterns (e.g. do you think the visualization is an accurate representation of your interests).

Participants were also guided to examine cross-platform data inference accuracy. They were encouraged to look at the different color-coded semantic clusters to identify patterns not known before, and to think if this could be an accurate representation of their interests on each platform. When the study was carried out in pairs or groups, we sometimes invited participants to make an inference of their peers' social media interests based on the visualization of their mirror, if the group was sufficiently comfortable and already familiar with each other.

\subsubsection{Part 2: Temporal Exploration of Mirror (15 minutes)}

After the open exploration, researchers guided participants to observe the temporal analysis on their Mirror, by either playing a time-lapse overview provided by our tool or zooming into a specific month/day on their timeline. Similar to part 1, participants were then given a chance to freely explore their temporal data while thinking aloud during their explorations. The exploration was guided by open questions to participants, e.g. whether seeing data chronologically gave them a different understanding of themselves, or whether they noticed any patterns or shifts in their behavior that they were not aware of before. Researchers took observations and notes throughout. 

\subsubsection{Part 3: Closing Interviews (10 minutes)}
At the end of the study, we conducted semi-structured interviews in individual or small group formats. The questions were designed to examine (a) participants' overall experience with the Mirror (e.g. do you feel you learned something new today); (b) their perception of datafication and inferences (e.g. has the tool changed the way you think about how platforms collect your data and influence what you see online); (c) their sense of data control (e.g. has it changed what you may want to do with your data on social media platforms?); and (d) any future features they would like to see. As above, researchers would always invite participants to expand on their responses by asking them to explain what they meant when they mentioned things like `cookies' or `business model' and by prompting them to expand on the `why's of their thoughts. Previous research~\cite{wang2022don} has demonstrated that such follow-up questions are crucial for researchers to understand what younger participants actually mean when they refer to abstract terms, and effectively dive into their rationales, values, and motivations.


\subsection{Data Collection and Analysis}
All sessions were video-recorded with participant and parental consent, following the IRB approval, yielding approximately 949 minutes of data. Screen recordings captured participant interactions with the visualization tool to provide behavioral context for verbal responses.

We employed thematic analysis following established qualitative research practices in HCI. Two researchers independently coded 20\% of the transcribed data to develop a preliminary codebook, followed by team discussions to resolve disagreements and refine codes. We note that, while our findings highlighted some themes related to participants' emotional expressions, we did not apply sentiment analysis, which may endanger over-interpretation of our data given the short time span of the study. These themes largely emerged from the explicit expressions from the participants. Observational notes taken during the interactive sessions were integrated with interview transcripts to provide comprehensive understanding of participant responses and behaviors.

\section{Findings}\label{sec:findings}
To address our research questions, we first outline participant adolescents' existing experiences and perceptions about datafication on social media platforms (\textbf{RQ1}). Next, we discuss their user experience with the Algorithmic Mirror and how it influenced their perception of datafication (\textbf{RQ2}). We then examine how Algorithmic Mirror affected their reflection of their online experiences being shaped by inferences (\textbf{RQ3}). Finally, we explore their needs for gaining more autonomy over their data and any barriers (\textbf{RQ4}). Participant quotes are presented with their ID and age for context.

\subsection{Existing Use of Social Media Platforms and Awareness of Datafication}\label{sec:existing}

Adolescents' awareness of datafication is largely consistent with literature, while their awareness of cross-platform datafication and how data can be tracked over time is particularly limited. Participants were largely aware of the use of \textit{recommendations} by algorithms on these platforms. Some explicitly linked their current recommendations to past watch history (\textit{``Well, the algorithm uses the [watch history] data to recommend me similar content to what I watch''} [P7, age 14]), while others described recommendations as a function of their likes or interests (\textit{``it recommends me stuff I think it thinks I would like''} [P2, age 14]; \textit{``I guess the algorithm itself makes sense, like, if you like this, we'll recommend more of it to you''} [P4, age 12]). 

However, awareness of cross-platform data sharing was particularly limited. A handful of participants thought data could be shared across platforms but only if the accounts were linked by the user (\textit{``[...] but it's not necessarily connected to my Instagram account, though not necessarily, but it will do basically mostly the same thing if you connect''} [P7, age 14]). A few participants mentioned `cookies' as a way that companies could track them across platforms, but none linked this cross-platform datafication to their recommendations (\textit{``I thought it was just for one specific platform. They'd take the information and then use that to recommend more''} [P5, age 13]). While some adolescents mentioned their awareness that companies aim for users to spend more time on the platform, they struggled to make sense why platforms that are supposedly competitors or for different purposes would share data (\textit{``It doesn't really make sense to me because why are you tracking my activity on an irrelevant app?''} [P26, age 14], \textit{``it's cool that even though they're rivals, they still share this data''}, [P2, age 14]).

Similarly, participants had hardly thought about how their data may be tracked and collected by platforms over time. When prompted, a few participants guessed platforms retained their data forever. Others thought data would only be retained for a couple of years as it would quickly become irrelevant: \textit{``this time last year I wasn't watching the same things on TikTok as I am now, so it's like, what would be the point in it saving the data from last year as well?''} [P26, age 14]. The temporal visualization of several years of social media watch history surprised many participants and led to immediate excitement, surprises and deep reflections.

\subsection{Experience of Datafication and Algorithmic Inference}~\label{sec:datafication}
After opening their individual Algorithmic Mirror, which includes a semantic clustering visualization from their watch histories, participants were given the opportunity to explore it at their own pace. We encouraged them to make better sense of inferences made about them by zooming in/out, while verbalizing their experiences by talking aloud. Seeing their personal data visualized often provoked a range of emotions, often beginning with an initial sense of excitement and surprise to feelings of nostalgia. The surprises were, e.g., associated with how accurate Algorithmic Mirror captured their interests or how different they thought their actual interests were. They were also caused, e.g., by the sheer number of videos they had watched over the years. These very personal connections with the visualizations played a crucial role for participants to zoom in, examine closely their watch behaviors, and reflect on how these data are being collected, processed, and used to make inferences about them.

\subsubsection{Sense of data collection}

After engaging with reflective interventions with the Mirror, almost all adolescents reported improved awareness of the amount of data being collected, especially across platforms or over time. For example, P1 described that \textit{``It's interesting because I knew the algorithm and that it was there and [..] influencing me, but I didn't realize that it could be taken and mirrored in such an extent''} [P1, age 14]. Some participants had data extending back to 2015, when they were 4 or 5 years old, and many participants reflected how they hadn't realized the amount of data that could be tracked by social media platform over years: \textit{``Did not expect that they can store data from 2023. It’s amazing to see how much data is tracked. I feel scared, [they are] obsessed to track my interests''} [P19, age 14]. The word `mirror' was often used by participants in the study to refer to how the visualizations made them feel, like \textit{``a copy of their childhood over the years''} [P20, age 14].

By looking at how their data could be collected by different platforms, participants became concerned that data were shared across platforms, and how this may affect the content they want to see, e.g., as said by P20, age 14, \textit{``when data is shared across platforms without my awareness. We are using specific platforms for specific purposes. I wouldn’t want my Netflix shared with [redacted]. Don’t want them to mix up. Will be hard to categorize the platform and then use them specifically.''}.

Others were concerned about cross-platform data sharing, as they feared that it could limit their ability to discover new content and pursue new interests on a different platform. This was reflected by P5, age 13, \textit{``If you're trying to get into something new or look for new interests, but it keeps recommending the ones that you had a while back or the interests that you looked into on a different platform.''}.

A small number of participants expressed willingness with cross-platform data sharing, recognizing the convenience of accurate personalized recommendations. They saw no harm in platforms accessing their data, even considering it as an exchange of platforms' revenue (\textit{``because there's lots of shows I watch like snippets on YouTube and then I see them (on Netflix) [...] And I'm like, which Netflix movie is this\textinterrobang\, So yeah, I would quite like to get recommended from all platforms''} [P1, age 14].

\subsubsection{Sense of data profiling and inference}
The majority of participants developed a better understanding of the mechanisms driving datafication and algorithmic inferences. They understood better how their data is used as inputs for these algorithmic processes, which ultimately shape content recommendation. For example, many of them were able to draw the connection between the content they watched and what possible inferences or profiling algorithms may have made about them: \textit{``Probably because I searched it up initially on like advice or related videos, but then after that the algorithm just suggested me more and more content as related to basketball''} [P16, age 14]. 

Based on this understanding, some of them were also able to critique how an algorithm may make inferences, e.g., \textit{``If you watch this for their shorts, it means that you're like this type of age though.''} [P7, age 14]; \textit{``I suppose you could conclude that the person that watched these things is probably based in either Asia or someplace where Japan has soft powers such as the US, maybe Europe to a lesser extent''} [P13, age 15]. 

The majority of the participant adolescents found the power of algorithmic inferences `exciting', `cool', and `very interesting', even though many of them did not like their data being collected, shared and processed in this way. In addition to personal interests, a few participant felt algorithms may be able to infer life events or moods based on content watched:

\begin{quote}
    \textit{``It's like a timeline essentially of what I've been watching, how I've been feeling, you know, like what's going on, you know? And it's, I think it's really interesting that you can see that, but it's also kind of concerning that you can learn so much about what's happened to me in the past six months from my TikTok usage''} [P27, age 14].
\end{quote}

At the same time, many of them expressed fear of surveillance and control, as noted by some participants below:
\begin{quote}
\textit{``It’s interesting but also scary, that without my awareness that social media can control what I am exposed to online. It’s really useful to reflect on what data to share''} [P18, age 14];

\textit{``I feel being watched over. It’s good to [..] they are doing their best''} [P19, age 14];

\textit{``I think any data that can allow like people to recognise you is a bit scary. Like for example, your age. I think YouTube can already kind of recognise that I'm a teenager with the things they give me so. Stuff like that. It's a bit scary of what they can do with that, but at the same time, like what kind of ads they're going to give me based on what I watch''} [P14, age 15].
\end{quote}

Furthermore, a small number of participants were willing to accept data processing and inferences, despite understanding well how these systems operate upon their data (e.g. \textit{``If I had to get recommended by a computer rather than a human, I'd rather it took data from all platforms that I (use)''} [P1, age 14]). They are reluctant to see changes as they enjoy the convenience and utility and they believe that this is their priority: \textit{``That you would find mostly the information that you really like to watch, for example. [...] I don't really feel anything to be like changed or anything like that. I actually be feel like feel like everything is like right though, like it's amazing.''} [P7, age 14].

\subsection{Reflection on Implications for their Identities and Online experiences}\label{sec:reflection}

\subsubsection{Reflection of self}

We observed that participants reflected in very different ways when using Algorithmic Mirror.
Many noted how Algorithmic Mirror enabled them to reflect deeply on the alignment between watch history and their personal interests: \textit{``Happy for platforms to collect the data and watch the videos. But also like to know the resources. Want to be comfortable''} [P19, age 14]. Many participants were impressed by the tool's ability to accurately reflect their viewing patterns, and the match between inferred interest and content they actually consumed: \textit{``It surprises me how well it's how well it's gotten, what I watch, how it's how well it's noted it down. [...] It's quite accurate, especially the gaming parts, because it's gotta correct down to the games that I play as well. So that's it's very detailed''} [P5, age 13]; \textit{``I like fast food. [...] I watch a lot of like video game videos as well. So I feel like that makes a lot of sense''} [P12, age 16].

However, some participants noted that the profile Algorithmic Mirror created from their watch history, and thus the algorithmic representation, was not an accurate representation of their true interests: \textit{``how much of it is just so irrelevant to what I do''}, \textit{``even if I don't remember watching these videos, they've happened and they've been collected and that's what the, the algorithm thinks I am.''} [P1, age 14]. A large number could not remember watching the videos listed (\textit{``Yeah, I can't remember watching half these videos.''} [P1, age 13]). This was particularly the case for adolescents who watch mostly short-form content, with limited descriptions that make it difficult for us to accurately categorize into clusters.

Ou researchers prompted them to reflect why the content they watch may not align with their interests, yet is recurrently shown to them by platforms (\textit{``Probably because I searched it up initially on like advice or related videos, but then after that the algorithm just suggested me more and more content as related to basketball.''}[P16, age 14]). 

Participants were surprised to find ads in their watch history, because they could rarely remember search for or even watching such content. This included, e.g., clusters of videos labeled `writing assistance' or `gift cards'. Some were surprised that inferences could be made about them based on advertisements watched, compared to `true' content. They quickly understood the situation once they looked more closely:
\textit{``Oh, I'm just thinking, if the YouTube history saves ads, does it? 
'Cause if it does, it will explain all the `writing assistance' with Grammarly ads.''} [P2, age 14].

The temporal visualization prompted some participants to reflect on their changing interests over time, which they sometimes depicted as `phases'. Even though the viewing had taken place several years ago, a large proportion of participants could recall their mood (\textit{``I only watched one episode of this just for like the nostalgia because I used to watch it when I was a child''} [P5, age 13]), social setting, and context that motivated what they had watched (\textit{``The Minecraft movie was out. So around April there's a lot of Minecraft content developing''} [P14, age 15]), as well as why they had stopped watching certain topics (\textit{``I grew out of it because it was a bit hard since I live in [redacted], most of them were based in other parts of the world, so they'd usually be streaming at late nights for me and I found it inconvenient''} [P5, age 13]). 

\subsubsection{Reflection on long-term impact}

A few participants expressed frustrated at the way their attention span had been affected by their use of platforms. Some noted  that they spent more time on the platforms than they wanted to: \textit{``sometimes when I'm using TikTok I get very carried away and I can spend more time on it than I intended to.''} [P6, age 17]. We found that such discussions took place more in the context of watching short videos than long videos on YouTube or Netflix.

A number of participants further spoke about the impact of their past viewing habits on their current self, describing it as a shaping force in their personal growth:
\begin{quote}\textit{``I feel like my online consumption often plays a role in character development as well.''} [P27, age 13];

\textit{``It's nice to see a reflection of what you've watched and how it's created the person you are now, watching yourself kind of evolve in a way.''} [P22, age 15].
\end{quote}

\subsubsection{Reflection on peer pressure}

Occasionally, participants discussed how their viewing habits were influenced by peers: \textit{``My taste has been influenced a lot by what girls my age like to watch, especially my friends at school [...].  But before [...] it was more my own personal taste.''} [P1, age 14]; \textit{``I switched schools [aged 13], so I was surrounded by different people and different things that they found funny.''} [P26, age 14].

\subsubsection{Reflection on inferences}

Many participants mentioned how the visualization helped them understand  how algorithms could infer information about them, especially over a long period of time. For example, the visualization made P24, age 15, see how they are perceived by algorithms and how their identity changed over their childhood, prompting reflections on datafication by algorithms:
\begin{quote} 
    \textit{``I feel like they captured my timeline of what my identity was from a younger age to an older age, which I think it's really fascinating because I I didn't realize that When I was younger, I mean, I still do not [...] I I guess I really liked Comedy too. I don't think I watched that much comedy recently. I don't think, but things like detectives, too. I think I used to. I think [...] it's really fascinating how they captured all my different likes [..] simple action can be recorded for a long time.''}
\end{quote}

The interviews showed that many adolescents could connect their digital footprints to how algorithms inferred personal traits about them, such as region, age, or gender identity, e.g. \textit{``I suppose you could conclude that the person that watches these things is probably based in either Asia or someplace where Japan has soft powers such as the US, maybe Europe to a lesser extent, etc.''} [P13, age 15].

Some were surprised by the level of details in the clustering of their interests made by Algorithmic Mirror, including e.g. specific games they played and videos they watched: \textit{``I was already aware, but I did not know how much of it they knew about me; I didn't think YouTube would mark every single video.''} [P11, age 15].

The data of some participants included a large number of very short videos, often called shorts or reels, for which Algorithmic Mirror showed a very dense visualization. These participants often reflected on differences observed across platforms. Some of them believe that they would be innaccurately profiled by TikTok's inferences, given they they were watching shorts mindlessly without noticing time spent on social media (e.g. \textit{``feeling spending too much time on screen time seeing TikTok visualization''} [P20, age 14]; \textit{``I think sometimes when I'm using TikTok I get very carried away and I can spend more time on it than I intended to''} [P6, age 13]); whereas others were more worried that TikTok may know more about them given that they could have so much more data about them than other platforms, even just over a few months (e.g. \textit{``I'm very confused of the piano thing. I think I like music a lot, but like I don't like piano, so maybe they thought that I would like piano. Yeah. I mean, it's kinda incorrect. Like I don't like.''} [P8, age 12]). The participants' perception of algorithmic inferences varied across platforms, and this was reflected in their discussing on coping strategies (see below).

\subsection{Desired Future for Data Autonomy}\label{sec:future}

Participants described several strategies they had devised to \textit{``influence''} algorithms, by \textit{``search[ing] for specific videos''} [P15, age 15], \textit{``watching more of one genre''} [P16, age 14], or \textit{``[not interacting] with them''} [P10, age 13]. However, they welcomed tools such as Algorithmic Mirror and wished such mechanisms could be \textit{``integrated in the systems''} [P20, age 14], demanding better support in future designs. In parallel, they expressed concerns about barriers for providing access and exercising controls, and fears that it would prevent platforms from providing exciting new innovations for them.

\subsubsection{Transparency mechanisms}
Several participants demanded better transparency regarding the use of their data. They wanted to be able to access and control this information themselves, including inferences from them (\textit{``I feel like they that they should let you see your own data and they should let you control it and look around in it}. [P3, age 16]); \textit{``If they are collecting your data, they are very transparent about it and they can let you just click maybe the three dots at the top of the page and then.  There'd be a button, say data or something like that, so you don't have to dig really deep in order to find what you've been watching and what the app has saved about you''}. [P1, age 13]).

Others participants expressed that being aware of data collection and inferences would help them feel more comfortable and in control, even if they would be unable to take further control of their data: 
\textit{``I could make myself more self-aware, but I have no say in what they're doing and what they collect from what I watch. [...] It wouldn't change anything for them, but it's nicer for me because if they're going to collect data about you anyway, then you can know, like, to what extent''} [P6, age 13].

However, a few participants voiced concerns, or dilemma, about the prospect of being unable to make effective changes: \textit{``I think that at times it [control] could be useful, but it also it's also kind of like if you're not aware of it and if you're not comfortable with it, it's not very, it's not a nice thing to know that. 
Everybody's aware of what you're interested in. It's like a pair of digital eyes spying on you.''} [P5, age 13].

\subsubsection{Control mechanisms}

Overall, participants often felt that they lacked control over their data, its use, and its interpretations by platforms (\textit{``sometimes I want to find something else...I wanna find something out of that box; I still rather if they give me like a more wider range''} [P9, age 15]; \textit{``I don't because they do take lots of information about me, but I can't really control that. So then I can't really do anything''} [P8, age 12]).

Some participants described specific control mechanisms that they would like to have, such as the options to erase their history. P1, age 14, suggested that the option should always be available and force algorithms to forget all past inferences:  \textit{``that means the algorithm will then know nothing about you from then on until you give it more data.''}. However, others feared potential trade-offs, such as losing access to some features, and wanted to know the consequences of deleting their data.

Many found it challenging to accept trade-offs required  to forgo access to context, as described by P3, age 16: 
\begin{quote}
\textit{``They they have the data now so like I can't really do anything about that because I can do something so they don't get as much data from now on. But it's kind of hard assuming that like I want to still keep watching YouTube and Netflix, I don't want to just like stop now.''}
\end{quote}

Others expressed concerns about the potential complexity involved in being responsible for their own data: \textit{``There's there's so many data points that it's almost impossible to just like cherry pick certain data points that you don't like people to know''} [P15, age 15]; \textit{````But the Netflix especially wait, did you not notice how buried it was in order to get our data like we had to click on a load of buttons before it showed us the data?  Quite interesting''} [P1, age 13].

Teenagers' desire for increased control over their data stems from a concern that data is being shared without their knowledge, e.g. \textit{``when the data collected to about me is shared between platforms without my awareness. Some people may use a specific platform for specific purposes and that being shared with your other applications can be unwanted sometimes, unable to distinguish content between different platforms''} [P20, age 14]. They may also want to control how their personality is perceived by platforms (\textit{``I watched a few videos on performance pressure in the K pop in the street. So, on the surface, it can be like oh I'm nervous about an upcoming performance. But in reality, it's something different.''} [P22, age 15]). A few remain pessimistic about alternative data ownership models, believing that data is key for innovation (\textit{``I understand that some people wouldn't like to like websites to collect your data but the reality is that's how it works, you can't change it''} [P19, age 14]).

\section{Discussion}

We proposed a new way to make datafication and algorithmic inferences visible to adolescents by providing personalized visualizations of their own data. Our contributions are threefold: (1) providing adolescents with a clear understanding of the scale and reach of data collection across platforms over time, (2) empowering them to establish a personal connection with their own, past experiences by visualizing their data, (3) fostering self-reflection by presenting them how algorithms may represent them, which can strengthen their motivation to reclaim digital autonomy. 

These provide important insights for future design directions, by stressing the importance of fostering adolescents' connection with lived digital experiences as an additional means to nurture their agency development. It also strengthens the demand for changes by social media platforms, to empower adolescents with more transparency of datafication, and more importantly, the agency to control how they are datafied and influenced by algorithms.

\subsection{Current Adolescents' Awareness of Datafication and Inferences}

By observing how adolescents perceive algorithmic inferences, our study reveals important gaps in their understanding and agency. This confirms findings from prior literature that highlight teens' limited awareness of algorithmic manipulation and their struggle to exercise meaningful control over personalized content systems. We show that teens largely believe that algorithmic inferences are designed to provide utility or convenience for them~\cite{ofcom2023,livingstone2018european,swart_experiencing_2021}. Furthermore, more than half of our participants had trouble making sense of how inferences or recommendations are carried out on short-video platforms such as TikTok or YouTube Shorts. This contrasts with previous studies with adults, who often exhibited nuanced strategies for influencing these algorithms according to their interests or perceptions of themselves~\cite{lee2022algorithmiccrystal}. These findings show a gap in teens' knowledge. The experience of continuous scrolling gives them the impression that little control can be applied or is effective, which often leads to a sense of despair or losing control. This is consistent with prior studies~\cite{5Rights2023, sahebi2022social, peterson2020negotiated, furnham2019automation}, which have shown that children and adolescents can be easily affected by the personalized manipulations exercised by such social media platforms without realizing the manipulative effects. 

Our study also shows that adolescents largely struggle to make sense of cross-platform datafication and have limited awareness of the scale of datafication over time, aligning with previous findings~\cite{livingstone2018european,smahel2020eukidsonline}. Algorithmic Mirror helped participants develop a better understanding of datafication at scale by providing personalized visualizations of inferred identities across platforms and time. Adolescents' reflections tended to focus more on their self identities and the content they consumed (see \Cref{sec:future}), rather than on the personal data itself being datafied by platforms. Nevertheless, our findings point to promising directions for understanding how teens perceive datafication on social media and how their values of their digital selves directly influence their choice of data autonomy. 

\subsection{Algorithmic Mirror: Enhancing Future Self-Reflection on Digital Identities}



Algorithmic Mirror provides a unique, contextualized visualization based on a user's real social media datasets. This approach of visualizing real users' data is also referred to as a `situated' visualization, and this use of real data to affect children's awareness of datafication reflects Piaget's pioneering theory of constructivism~\cite{piaget1972development}, which holds that children learn more effectively by actively constructing knowledge through their own experiences. Our situated visualization supports Papert's theory of constructionism (1991), which argues that when learning environments are grounded in personally meaningful contexts, learners can bring their own lived experiences into the educational process, enabling deep reflection and understanding that abstract simulations cannot replicate~\cite{papert1991constructionism,dignazio_creative_2022,dangol_constructionist_2023}. 
It also incorporates key aspects of reflective informatics: breakdown, inquiry, and transformation~\cite{baumer_reflective_2015}.

Our use of situated visualization enabled participants to personally connect with their visualization via an open exploration process.
This was evident from their excitement and extensive interest in exploring and examining  content deeply. This, in turn, prompted reflections about themselves and their online activities, and how these are shaped by algorithms, overall developing a better sense of the implications of datafication and inference (see \Cref{sec:reflection}). Below, we identify design features of Algorithmic Mirror that enabled the situated visualization and development of awareness. We highlight three interrelated mechanisms in our Algorithmic Mirror (enabling emotional connection through personal data, revealing the scale and scope of datafication, and supporting open exploration) that translated theoretical principles of constructivist and constructionist learning into concrete interactive affordances. These features anchored participants’ reflections in their own lived data, and  revealed the scale, continuity, and interconnectedness of datafication in ways that abstract explanations could not achieve.

\subsubsection{Enable Emotional Connection Through Personal Data}
When designing Algorithmic Mirror, we made the deliberate choice of providing the facility for adolescents to upload their real social media watching history to the tool. Making sense of up to raw watching history data, here totalling up to 60,000 videos, is challenging. By using interpretable clusters of videos, we constructing a meaningful narrative for participants to see how their online activities are perceived by algorithms (see details in \Cref{fig:pre_processing_pipeline}). This approach builds on previous research emphasizing the importance of embedding personal data within its contextual framework to enable meaningful self-reflection~\cite{carpendale_subjectivity_2017, ferreira_interactions_2023, lupi_dear_2016, mementos2016, kim_dataselfie_2019, liang_sleepexplorer_2016, lupton_self-tracking_2014, slovak_reflective_2017}. By exposing similar types of semantic relationships that platforms may use to make inferences about user preferences and drive content recommendations, Algorithmic Mirror enabled participants to `see' vividly how algorithms draw explicit connections between the content they watched and make inferences about them based on these behaviors (see \Cref{sec:datafication}). 
When a participant's experience and expectations do not match with what is shown by the visualization, this can lead to moments of breakdown---puzzling, surprising, or anomalous situations~\cite{baumer_reflective_2015}. The personal relevance of the data adds an additional emotional connection to the visualization that motivates participants to inquire further in an attempt to explain the anomaly, thereby creating the foundation necessary for deep reflection and critical awareness.

\subsubsection{Show Scale and Scope of Datafication}
Algorithmic Mirror reveals the true scale and scope of datafication by making visible both its temporal persistence and spatial reach across platforms. These two dimensions often remain hidden in users' everyday platform interactions.

\textbf{Temporal Dimension}: The use of temporal visualization (see  \Cref{fig:zoom_and_pan_and_temporal_evolution}) enabled participants to see datafication as a continuous activity. Before using Algorithmic Mirror, participants believed their data was stored only briefly (see \Cref{sec:existing}). Participants expressed genuine surprise at the scope and persistence of data accumulation, particularly discovering that platforms retained viewing histories spanning years, e.g. dating back to early childhood when they were four to five years old. Many expressed fascination with the evolution of these categorizations, how life transitions were reflected in their viewing patterns over time, which often builds up the crucial incentive for reflecting on how they would like their interests to be perceived. This contrasts with previous tools visualizing a single snapshot of data, failing to capture how platforms accumulate data over extended periods~\cite{eslami2015feedvis, bhargava2019gobo, mozilla2013lightbeam}. The temporal dimension offers adolescents the opportunity to grasp how algorithmic systems construct users' identity profiles and how datafication operates as continuous surveillance rather than occasional snapshots. 

\textbf{Spatial Dimension}: The cross-platform view (see \Cref{fig:datafication}) enables adolescents to conceptualize how datafication takes place across multiple platforms. This contrasts with previous tools visualizing data flows from the server side~\cite{datatrack2015,bier_privacyinsight_2016,hansen_designing_2018}, or displaying data from a single platform~\cite{wang2024chaitok, elkattan2023algorithm}. Ensuring that similar content content from YouTube, Netflix, and TikTok are visualized in the same cluster or area is a challenging task . To enable better cross-platform clustering, we implemented a preprocessing step that standardize content summaries (see details in \Cref{fig:pre_processing_pipeline}). The visualization of unified cross-platform data consistently surprised participants, who had previously compartmentalized their digital activities by platform. They could see, for the first time, how they can be collectively tracked and profiled across platforms.

\subsubsection{Let Teens Explore Freely}
For self-reflection, our findings also indicate a need to shift from designing for \textit{explanations} of algorithms~\cite{ehsan_human-centered_2022, maalvikaXAI, dhanorkar2021who} toward designing \textit{explorable mirrors}, i.e. systems that support epistemological pluralism by allowing personally meaningful investigative pathways~\cite{epistemological1990}. Our tool allowed participants to flexibly explore data by zooming between details and overviews, navigating to specific time periods, and filtering platforms through color-coded selections. This inquiry-based approach aligns with emerging work in algorithmic auditing that similarly emphasizes participatory investigation and critical questioning of algorithmic systems~\cite{morales2024youth, morales-navarro_learning_2025, solyst_potential_2023}, as well as that of reflective informatics, that uses moments of breakdown as an opportunity to engage in conscious, intentional re-examination of currently held ideas, theories, and concepts~\cite{baumer_reflective_2015}. Such a turn would move HCI away from treating algorithmic literacy as a matter of comprehension and toward treating it as \textit{inquiry}: a space where users, especially young ones, can ask their own questions, develop new understandings of their experiences in digital systems, and test their developing sense of agency~\cite{piaget1972development, pedaste_phases_2015, morales-navarro_conceptualizing_2023}. 
However, free exploration does not necessarily mean an absence of scaffolding, and indeed \citet{fleck_reflecting_2010} highlight the importance of the right environment for reflection in the form of time, suitable structure, and encouragement.
Without suitable guidance and expertise to help adolescents make sense of what they are exploring (as was provided by the researchers in this study), the tool might not be as effective~\cite{slovak_reflective_2017}.

In summary, building on adolescents' strong personal and emotional connection with their data, Algorithmic Mirror provides a promising scaffolding tool for teens to develop critical awareness of datafication in personally meaningful contexts. We saw that this situated reflection extends beyond localized interpersonal privacy tactics (such as adjusting simple account settings), toward an awareness of \textit{systemic data practices}, which is foundational for children to develop a more critical stance at datafication by recognizing its broader, systematic impact~\cite{livingstone2018european,smahel2020eukidsonline}. 
This approach of setting personal contexts contrasts sharply with previous research on adolescents and datafication using simulated data~\cite{wang2024chaitok, elkattan2023algorithm}. While previous work did improve the understanding of the abstract concept of datafication, Algorithmic Mirror enabled adolescents to construct a \textit{situated} understanding of how algorithms personally affected them. 

\subsection{Adolescents' Demands for Agency}
\textit{Algorithmic Mirror} enabled adolescents to reflect not only on their self-perception but also on how algorithms construct their identities across platforms and over time. This is especially important given that adolescence is a crucial stage of identity development, negotiating questions about authenticity, boundaries, and continuity while experimenting with different roles and self-presentations. Our findings confirm that adolescents care deeply about how they are represented and perceived in algorithmic systems (just like in the physical world) and \textbf{demand agency in the categories and parameters assigned to them}. This agency is rarely recognized in existing social media platforms, and our findings provide new design opportunities both for the HCI community and the social media providers. 

Some participants expressed openness to datafication; however, this practice is only acceptable when it produces a profile that is \textbf{authentic} and true to their real interests. They regard distorted categorizations as threatening their sense of authenticity, while permitting inferences and datafication, as long as they are aligned with their intentions and interests. 

\textbf{Boundaries} are another paramount desire for teens---keeping a strict limit around what parts of themselves should remain private, especially in sensitive domains such as mental health or body image. They are not only concerned that this is intrusive to their personal space but also how such inferences could enable predatory advertising targeting their vulnerabilities, showing critical engagement with the risks of algorithmic profiling and identity protection.

Seeing persistent categorizations from childhood revealed a clash between developmental fluidity and algorithmic permanence. Adolescents are keen to preserve their developmental past as part of their `self-identity' and desire for the \textbf{agency} to control how algorithmic systems reflect their `current' preferences, by ``resetting'' or reweighting profiles to align, reflecting Erikson’s idea of adolescence as a time of identity exploration and self-authorship~\cite{erikson_identity_1980, sokol_identity_2009}.

Similarly, teens desire the \textbf{agency to maintain distinct identities} across platforms, resisting the homogenization of their interests into a single profile. For each social media site, they have a slightly different presentation of self~\cite{hogan_presentation_2010}. They valued platform-specific identities—Netflix for certain genres, TikTok for others—and wanted to control how these fragmented selves were stitched together. 

Our findings show that \textit{Algorithmic Mirror} directly supported adolescents’ developmental need for negotiating identity by giving them visibility into algorithmic categorizations and empowering them to make better sense of their digital selves. As a result, their motivation to act did not stem from abstract calls to ``take back control of data,'' but from personal concerns about how they were seen and represented, whether as authentic, controlled, evolving, or distinctive selves. This self-motivated reflection and action-taking respond to teens' inner needs and align with self-determination theory, which highlights that inner motivation is crucial to the development of autonomy for young people to learn to set goals, monitor progress, and adjust strategies.

By surfacing their digital identity in ways that resonated with adolescents’ immediate developmental needs, Algorithmic Mirror transformed abstract notions of datafication into a personally meaningful reflection and a process of self-motivated identity confirmation. This demonstrates that designing for \textbf{identity relevance}, not just data literacy, may be the key to supporting adolescents’ data autonomy. Our findings also points to the need to shift from building “reflection tools” that help users see data, to designing \textbf{agency-supporting systems} that scaffold iterative identity reflection and confirmation. This could transform the focus of adolescents algorithmic literacy research from just awareness-raising to self-regulated learning, encouraging the development of intrinsic motivation and capability building.

\subsection{Design Implications for Supporting Sustained Behavioral Change} 

Our findings reveal three key design implications for transforming personal data visualization tools from one-time research instruments into sustained interventions that could support meaningful behavioral change over time~\cite{reflection2010}. Similar to how a mirror provides immediate self-reflection and intentional self-presentation, these future visualization tools should serve as lasting partners throughout adolescent development, continuously supporting their digital agency. Effective systems should bridge the gap between reflection and action, operate dynamically rather than as static snapshots, and provide users with agency over how their digital identity is represented and explored.

\subsubsection{From Reflection to Action}
Participants frequently transitioned from passive observation to active speculation about their `ideal online self', articulating specific strategies for influencing algorithmic inferences. This behavioral shift indicates that effective personal data tools should support the metacognitive processes central to self-regulatory learning~\cite{Zimmerman2002}. Future systems could support users in constructing complementary maps: one representing `where they are now' based on actual data patterns, and another representing `where they want to be' configured based on their values and goals. This approach aligns with speculative design methods that create alternative narratives and envision potential futures, challenging current assumptions while fostering imagination and revealing new possibilities~\cite{wakkary_material_2015, candy_designing_2017, forlano_speculative_2023}. Such comparative mapping would enable adolescents to visualize the gap between their current and ideal mirrors, providing actionable insights that bridge reflection and action.

By examining differences between these representations while reflecting on their motivations for change, participants can identify specific steps---engaging with different content types, following new creators, or changing interaction patterns---to transform their digital consumption toward their aspirational self. However, translating these insights into sustained behavioral change requires careful consideration of how such tools will be implemented beyond research contexts.

To preserve the authentic self-reflection observed in our study, future systems should prioritize adolescent agency of these actions over adult supervision. Rather than serving as monitoring tools for parents or assessment instruments for educators, these systems should function as personal development resources that adolescents control directly. While independent use eliminates researcher guidance, systems could incorporate embedded scaffolding mechanisms. Scaffolded reflection prompts could help users interpret their data patterns and identify misalignments between current and aspirational selves, while micro-interventions---such as periodic check-ins about goal progress or contextual nudges when consumption patterns diverge from stated values---could provide structured support for translating insights into concrete action plans. Future systems should bridge reflection and action by offering these integrated guidance tools that support self-regulatory learning processes, enabling participants to think deeply about their motivations for digital consumption and inspiring the informed agency that leads to meaningful behavioral transformation.


\subsubsection{Dynamic and Real Time Systems}

Participants particularly appreciated visualizing their data over time. They expressed strong interest in real time or continuously updated versions of Algorithmic Mirror, inline with research on virtual body ownership demonstrating that synchronous mirror reflections create stronger subjective identification with one's avatar compared to asynchronous reflections~\cite{Gonzalez2010VRreflections}. When extending this work to longitudinal studies and behavioral interventions, personal data tools should operate continuously rather than as one-time interventions. Ensuring that participants can visualize recent data is, however, difficult. The static nature of data exports provided by platforms requires changes to our technical architecture. Implementing such continuously updated systems would require sustained API access that current platforms do not support, and regulations do not request.

\begin{figure*}[t]
    \centering
    \includegraphics[width=1.0\textwidth]{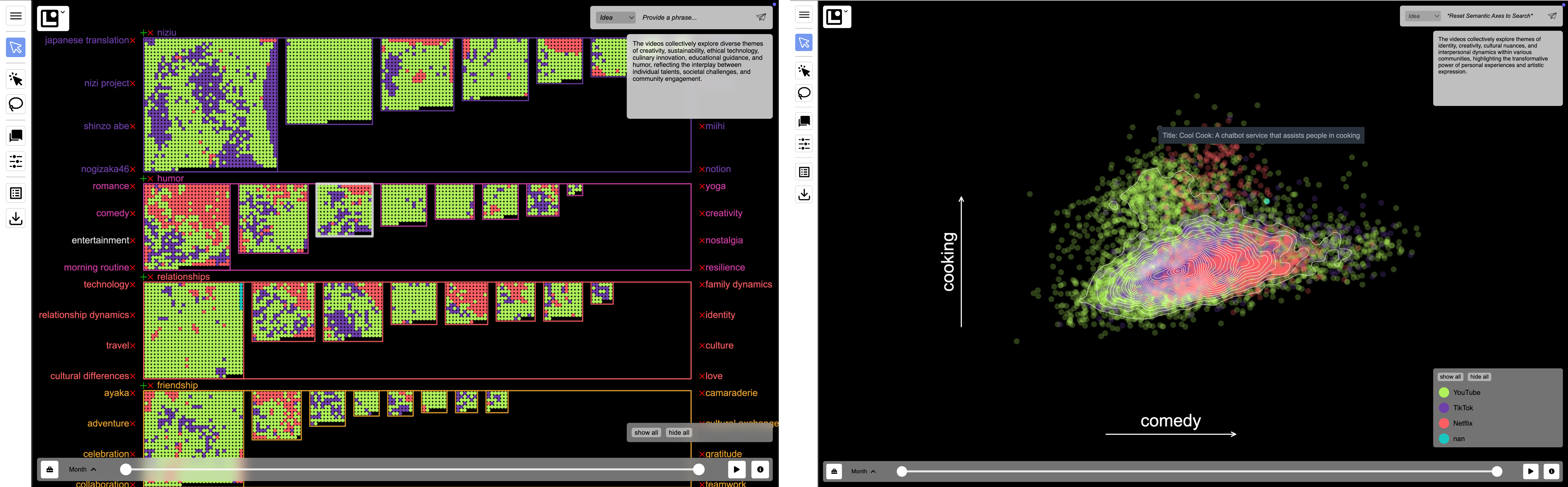}
    \caption{Grid and Semantic Axes layouts. These experimental configurations were tested with a small group of participants interested in greater control over the Mirror’s organization. The grid layout (left) arranges videos by major topics and their subtopics in a structured grid, which updates dynamically as topics are added or removed. The semantic axes layout (right) allows users to define x- and y-axis concepts; each video is positioned by its similarity to these concepts. This view also supports identity plots of a single concept, as well as mapping the x-axis to time to see how concepts trend over time in their watch history.}
    \label{fig:grid_and_semantic_layout}
\end{figure*}

\subsubsection{User Agency in Data Representation}
Participants demonstrated a strong desire for agency over both their entry point into the visualization and how they could manipulate and reconfigure their data representation to align with their personal context. When shown alternative experimental visualization modes---including grid layouts and semantic axes (see \Cref{fig:grid_and_semantic_layout}) that allowed reorganization based on user-defined concepts, all participants expressed positive reactions to controlling how their data was structured and displayed. As P27 noted, "\textit{I like this layout as well because you can edit it. I like this mixing of words with the visuals, because on the }[semantic cluster] \textit{visual, I liked the clusters, but missed the specific-ness of what }[each cluster]\textit{ is.}"

This desire for control reflects the need to support self-regulatory learning, a process by which individuals develop awareness of their motivations and strategies for achieving goals. Users who reorganized the mirror by removing categories, adding new ones, or restructuring layouts actively speculated about their online identity. They examined not just what their data shows but why they consume certain content and how it aligns with their values.

\subsection{Implications for Policy and Future Platform Design}
Our findings confirm that adolescents are left behind a `one-way mirror', unable to reflect on the digital selves and profiles constructed by platform algorithms. Participants expressed deep concern about algorithmic perceptions and craved mechanisms to influence these profiles, yet felt they currently have few options to exercise control.

Specifically, harms lie not only in exposure to inappropriate content but also in systematically denying children agency over how they are profiled and perceived, while providing few opportunities for autonomy. Without affording adolescents the ability to manage or reshape constructed profiles, platforms can, e.g., deepen vulnerabilities around body image, mental health, and identity formation by amplifying related content.

If deployed on a national or international scale to combat such harms, our implementation of Algorithmic Mirror would face structural barriers. Platforms lack APIs for data enrichment, provide no programmatic export capabilities, and offer inconsistent data access. For our participants, some exports arrived within hours, others required weeks. The heterogeneity in data collection and formatting across platforms also underscores the need for standardization. While some platforms offer inference transparency, e.g. Google through My Ad Center~\cite{google2022adcenter} and Spotify through Spotify Wrapped~\cite{peng2022spotify}, comprehensive access for researchers and users remains limited. Legal challenges such as Zuckerman v. Meta underscore the unresolved tensions between platform opacity and users' rights to understand algorithmic processing~\cite{zuckerman2024lawsuit}.

Future policy frameworks must move beyond narrow privacy rights toward structural reforms that guarantee children both visibility into their algorithmic profiles and the agency to influence them. As Wachter argues, the right to reasonable inferences should encompass both transparency about profile construction and agency over how historical data shapes algorithmic treatment~\cite{wachterCounterfactualExplanationsOpening2017}. This requires mandating standardized data export formats, API access for personal data tools, and extending transparency requirements to include algorithmic inferences that shape users' online experiences.

\balance

\section{Limitations and Future Work}
\label{sec:limitations}

Several methodological limitations should be acknowledged when interpreting our findings. For participant samples, our recruitment was geographically skewed toward Japanese adolescents, with participants distributed across Japan ($n=16$), the UK ($n=4$), Europe ($n=4$), and other countries, including Bahrain and the US ($n=3$). This geographic distribution suggest that our tool is appropriate across a range of countries and continents, but limits generalization to other countries, e.g. in the Global South as well as different cultural contexts and privacy norms, particularly regarding how different cultures approach adolescents' privacy and autonomy. However, we have observed no distinct thematic differences across the geographical distribution.

While we collected nearly watch histories for nearly 750K videos from participants, we note limited cross-platform coverage, with only 30\% of participants providing data from multiple platforms. This substantially constrained our analysis of cross-platform datafication awareness---a core contribution of Algorithmic Mirror. Participants with multi-platform data had stronger reactions to a unified digital identity visualization. This suggests that broader platform coverage may yield even richer insights into how adolescents perceive their fragmented digital identities across services.

We could only assess short-term effects through our single 60-minute sessions, which captured immediate responses and stated intentions but could not assess longer-term behavioral changes or sustained engagement with data autonomy practices. While we observed differences in participants' immediate responses to their personal data visualizations, our study design did not focus on measuring actual behavior change outcomes. The gap between stated intentions and actual behavior change is particularly relevant for adolescents whose self-regulatory capacities are still developing, making it crucial for future longitudinal studies to examine the sustained impacts of personal data visualization on digital behavior patterns.


In summary, these limitations suggest important directions for future research, including longitudinal studies to test for sustained behavioral changes, increased sampling for richer insights, and cross-cultural studies that can accommodate the diverse nature of adolescent online experiences.

\section{Conclusion}

This research addresses the pressing issue of empowering adolescents with digital autonomy amidst pervasive datafication. We introduced Algorithmic Mirror, a personalized visualization tool that makes opaque profiling practices more transparent and allows adolescents to explore years of data collected by platforms, revealing how algorithms might perceive and categorize them across YouTube, TikTok, and Netflix over time. 

Our research, which involved user studies with 27 adolescents aged 12--16, revealed crucial design considerations for supporting teens' autonomy and agency on social media. These include: (1) providing adolescents with a clear understanding of the scale and reach of data collection across platforms over time, (2) empowering them to establish a personal connection with their own, past experiences by visualizing their data, (3) fostering self-reflection by presenting them how algorithms may represent them, which can strengthen their motivation to reclaim digital autonomy.

\newpage
\bibliographystyle{ACM-Reference-Format}
\bibliography{reference,10.allrefs}

\appendix
\section{Participant information}

\begin{table}[ht]
\definecolor{Gray}{gray}{0.9}
\centering
\caption{Participant demographics and which platform data they uploaded to Algorithmic Mirror. Where consecutive rows are shaded the same, the participants took part as a group, e.g., P9, P10, and P11 formed one group, P12 and P13 formed another group.}
\begin{tabular}{cccccc}
\toprule
ID & Gender & Age& TikTok & YouTube & Netflix\\
\midrule
P1 & F & 13 & $\times$ & \checkmark & \checkmark \\
\rowcolor{Gray}
P2 & M & 14 & \checkmark & \checkmark & \checkmark \\
\rowcolor{Gray}
P3 & M & 16 & $\times$ & $\times$ & \checkmark \\
P4 & M & 12 & $\times$ & \checkmark & $\times$ \\
\rowcolor{Gray}
P5 & M & 13 & $\times$ & \checkmark & \checkmark \\
P6 & F & 13 & $\times$ & \checkmark & \checkmark \\
\rowcolor{Gray}
P7 & F & 14 & $\times$ & \checkmark & $\times$ \\
P8 & F & 12 & \checkmark & \checkmark & \checkmark \\
\rowcolor{Gray}
P9 & F & 15 & $\times$ & $\times$ & \checkmark \\
\rowcolor{Gray}
P10& F & 13 & $\times$ & \checkmark & $\times$ \\
\rowcolor{Gray}
P11& F & 15 & $\times$ & \checkmark & \checkmark \\
P12& M & 16 & $\times$ & \checkmark & $\times$ \\
P13& M & 15 & $\times$ & \checkmark & $\times$ \\
\rowcolor{Gray}
P14& M & 15 & $\times$ & \checkmark & $\times$ \\
\rowcolor{Gray}
P15& M & 15 & $\times$ & \checkmark & $\times$ \\
P16& M & 14 & $\times$ & \checkmark & $\times$ \\
P17& M & 14 & $\times$ & \checkmark & $\times$ \\
\rowcolor{Gray}
P18& F & 14 & $\times$ & $\times$ & \checkmark \\
\rowcolor{Gray}
P19& M & 14 & $\times$ & \checkmark & $\times$ \\
\rowcolor{Gray}
P20& F & 14 & \checkmark & \checkmark & $\times$ \\
\rowcolor{Gray}
P21& M & 14 & $\times$ & \checkmark & $\times$ \\
P22& F & 15 & $\times$ & \checkmark & $\times$ \\
P23& F & 15 & $\times$ & \checkmark & $\times$ \\
P24& F & 15 & $\times$ & $\times$ & \checkmark \\
\rowcolor{Gray}
P25& M & 13 & $\times$ & \checkmark & \checkmark \\
P26& F & 14 & \checkmark & $\times$ & $\times$ \\
\rowcolor{Gray}
P27& F & 13 & \checkmark & \checkmark & $\times$ \\
\bottomrule
\end{tabular}
\label{tab:demographics}
\end{table}

\end{document}